\documentclass[prl,twocolumn,showpacs,amsmath,amssymb,superscriptaddress]{revtex4-1}
\usepackage{graphicx}
\usepackage{breqn}
\usepackage{xcolor}
\usepackage{color}
\usepackage{booktabs}
\usepackage{float}
\usepackage{longtable}
\usepackage{epsfig}
\usepackage{bm}
\usepackage{dcolumn}
\usepackage{lipsum, babel}
\usepackage{soul}
\usepackage{cancel}

\usepackage{rotating} % for sidewaystable

\usepackage[normalem]{ulem}
\usepackage{epstopdf}
\makeatletter
\let\cat@comma@active\@empty
\makeatother
\begin{document}

\title{P-wave magnets}

\author{Anna Birk Hellenes}
\affiliation{Institut f\"ur Physik, Johannes Gutenberg Universit\"at Mainz, D-55099 Mainz, Germany}

\author{Tom\'{a}\v{s}~Jungwirth}
\affiliation{Institute of Physics, Czech Academy of Sciences, Cukrovarnick\'a 10, 162 00, Praha 6, Czech Republic}
\affiliation{School of Physics and Astronomy, University of Nottingham, NG7 2RD, Nottingham, United Kingdom}

\author{Rodrigo Jaeschke-Ubiergo}
\affiliation{Institut f\"ur Physik, Johannes Gutenberg Universit\"at Mainz, D-55099 Mainz, Germany}

\author{Atasi Chakraborty}
\affiliation{Institut f\"ur Physik, Johannes Gutenberg Universit\"at Mainz, D-55099 Mainz, Germany}

\author{Jairo~Sinova}
\affiliation{Institut f\"ur Physik, Johannes Gutenberg Universit\"at Mainz, D-55099 Mainz, Germany}
\affiliation{Department of Physics, Texas A\& M University, College Station Texas 77843-4242, USA}

\author{Libor \v{S}mejkal}
\affiliation{Institut f\"ur Physik, Johannes Gutenberg Universit\"at Mainz, D-55099 Mainz, Germany}
\affiliation{Institute of Physics, Czech Academy of Sciences, Cukrovarnick\'a 10, 162 00, Praha 6, Czech Republic}
\affiliation{Max Planck Institute for the Physics of Complex Systems, N\"othnitzer Str. 38, 01187 Dresden, Germany}

\date{\today}

\begin{abstract}
The p-wave Cooper-pairing instability in superfluid $^{3}$He, characterized by a parity-breaking excitation gap, is regarded as one of the most rich and complex phenomena in physics. The possibility of a counterpart unconventional p-wave ordering of interacting fermions, in which a Fermi surface spontaneously breaks the parity symmetry, has been an open problem for many decades.  Here we identify the realization of the counterpart of p-wave superfluidity in magnetism. We demonstrate a strong parity-breaking and anisotropic symmetry lowering of spin-polarized and time-reversal symmetric Fermi surfaces in a representative p-wave magnet CeNiAsO.  As a direct experimental signature we predict a large spontaneous anisotropy of the resistivity. Abundant and robust realizations of the unconventional p-wave magnetism can be identified from suitable non-relativistic crystal-lattice and spin symmetries, without requiring strong correlations and extreme external conditions. This opens new prospects in fields ranging from topological phenomena to spintronics.

\end{abstract}

\maketitle

\section{Introduction}

The Fermi liquid theory \cite{Landau1957,Vignale2022} of elementary quasiparticle excitations near the Fermi surface has been one of the most successful frameworks in condensed-matter physics for describing many-body ordered phases of interacting fermions. 
In conventional s-wave superconductivity, the $\mathcal{U}(1)$ gauge symmetry is broken by the Cooper-pairing  instability of the Fermi surface, while preserving the crystal-lattice symmetry of the quasiparticle excitation gap formed in place of the Fermi surface \cite{Annett1995,Houzet2012,Tsuei2000,Mackenzie2003}. 
The unconventional p-wave Cooper pairing in $^{3}$He breaks the parity symmetry (Fig.~1(A)), in addition to the gauge symmetry \cite{Annett1995,Houzet2012,Tsuei2000,Mackenzie2003}. The resulting extraordinary properties make the  superfluidity in $^{3}$He \cite{Leggett1975} one of the most celebrated phenomena in physics, despite its occurence within  a limited phase space of very low temperatures and pressures.

In conventional itinerant ferromagnetism, the $\mathcal{SO}(3)$ spin-rotation symmetry is broken \cite{Moessner_Moore_2021} by an s-wave spin-dependent Pomeranchuk instability, while preserving the crystal-lattice symmetry of the spin-split Fermi surface, in analogy to conventional s-wave superconductors. 
A parity-breaking p-wave Pomeranchuk instability, originating from spin-independent interactions and resulting in a spontaneous charge current carried by the off-centered Fermi surface, has been explored since early attempts to explain superconductivity eight decades ago \cite{Born1948}. 
The possibility of the parity-breaking p-wave Pomeranchuk instability in strongly correlated systems has been theoretically considered for both the spin-independent and spin-dependent interactions \cite{Hirsch1990}. The constrains on the spontaneous charge and spin-currents in the ordered many-body ground state have remained an open problem, and  physical realizations of these parity-breaking phases have not been identified despite the eight decades of research \cite{Born1948,Bohm1949,Hirsch1990,Wu2007,Jung2015,Kiselev2017,Wu2018}.

Our identification of  unconventional p-wave magnetism (Fig.~1(B)), with an example of a robust realization in CeNiAsO, is based on a principally distinct approach sidestepping the unavailable exact solution of the many-body interaction problem. It follows in the  footsteps of the recent discovery of altermagnets with a counterpart even-parity-wave magnetic order to unconventional d-wave superconductivity \cite{Smejkal2022a}.  Instead of  focusing on strongly correlated Pomeranchuck instabilities, and limiting the search  to metallic electron fluids, the anisotropic symmetry-lowering of the spin-polarized energy isosurfaces of altermagnets was found to be promoted  by suitable  non-relativistic crystal-lattice and spin symmetries \cite{Smejkal2022a}.  Consequently, altermagnetism was theoretically identified over the entire range from insulators and semiconductors to metals and superconductors, often at ambient conditions \cite{Smejkal2022a}. Spectroscopic signatures of altermagnetism, predicted by spin-symmetry analysis  and density-functional-theory (DFT) calculations in MnTe\cite{Smejkal2022a}, have recently been reported in photoemission experiments detecting well-resolved spin-polarized quasiparticle bands \cite{Krempasky2024,Lee2024,Osumi2024,Hajlaoui2024}.

\begin{figure}[h!]
	\centering
	\includegraphics[width=1\linewidth]{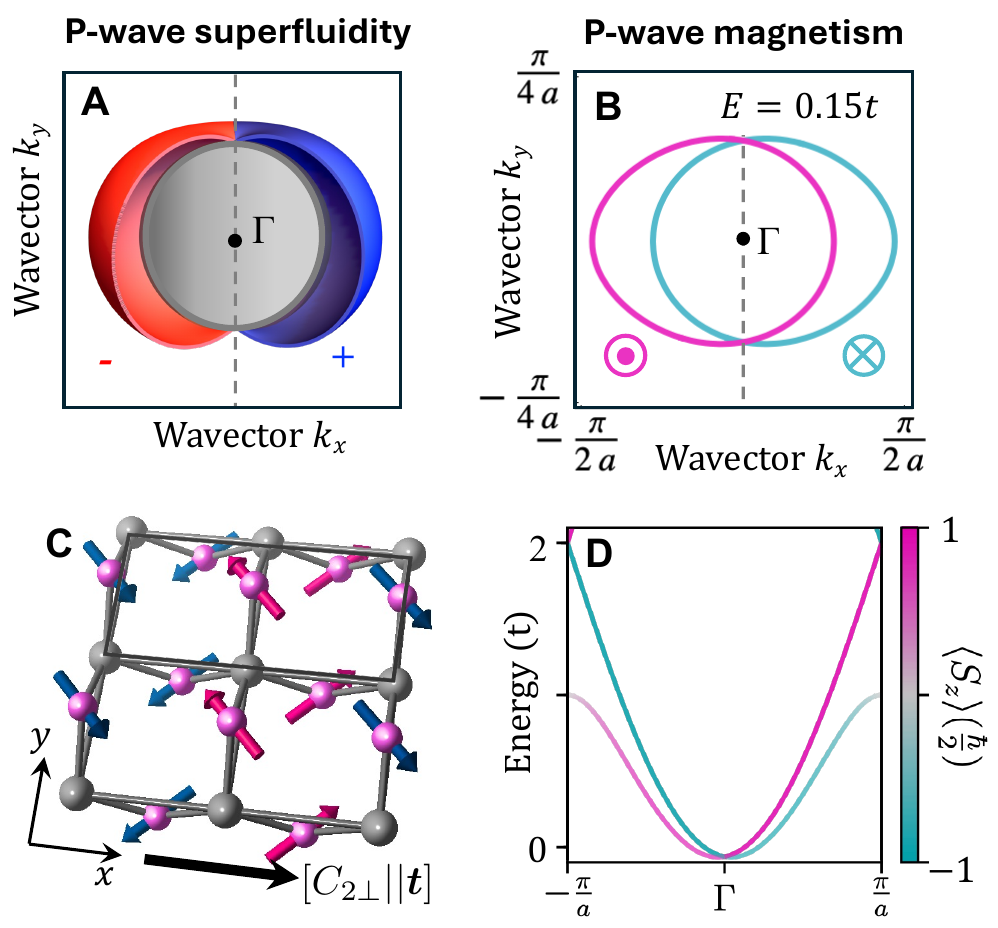}
	\caption{\textbf{Unconventional p-wave phases.}
(A) Cartoon of parity-breaking Cooper-pairing gap function in p-wave superfluid $^{3}$He. (B)   Spin-polarized Fermi surfaces of a model band structure of an unconventional p-wave magnet. The parity breaking is complemented by an anisotropic symmetry breaking of the Fermi surfaces. (C) Cartoon of a coplanar non-collinear spin arrangement on the crystal with the spin arrangement that realizes the unconventional p-wave phase. The spin arrangement breaks the inversion symmetry and exhibits the spin symmetry $[C_{2\perp}||\boldsymbol{t}]$. The corresponding  $[C_{2\perp}||E]$ symmetry in the momentum space protects the momentum-independent spin axis along the $C_{2\perp}$ spin rotation axis marked by out-of-plane arrow direction.
(D) Spin polarised energy bands corresponding to the model in panels (B,C). 
}
\label{fig:figure1}
\end{figure}

In this work  we identify  the non-relativistic crystal-lattice and spin symmetries favoring the unconventional p-wave magnetism, and confirm the symmetry principles by DFT calculations. 
In the p-wave magnet CeNiAsO discussed in detail below, we observe that the parity breaking of the spin-polarized energy isosurfaces is accompanied by an additional anisotropic distortion and that the spin polarization of these isosurfaces is collinear. In the following model and DFT calculations, we will focus on this p-wave order. We predict a large spontaneous anisotropy of the resistivity as a readily detectable experimental signature of this  unconventional magnetic phase.

Before presenting the DFT results, we first summarize the symmetry guidelines that led us to the identification of material candidates for the unconventional p-wave magnetism. Remarkably, to find materials with oppositely shifted off-centered energy isosurfaces (Fermi surfaces) with opposite collinear spin polarizations in the momentum space, we discard collinear magnetic orderings in the real-space crystal structure. This follows from the earlier complete classification of collinear magnets by the non-relativistic spin-symmetry groups showing that they only host even-parity band structures \cite{Smejkal2021a}. Instead, we focus on non-collinear coplanar non-centrosymmetric magnets with a non-relativistic spin symmetry $[C_{2\perp}||\boldsymbol{t}]$, where  $C_{2\perp}$ is a 180$^\circ$ spin-space rotation around the axis perpendicular to the plane of the coplanar magnetic moments in the crystal, and $\boldsymbol{t}$ is a lattice translation. (Throughout the text, the spin-space transformations are on the left of the double vertical bar and the real-space transformations on the right.)

The $[C_{2\perp}||\boldsymbol{t}]$ symmetry combined with the non-centrosymmetric magnetic crystal structure in real-space allows for a compensated magnetic phase with zero net magnetization, and a momentum-space energy spectrum, satisfying parity-breaking:
\begin{equation}
E({\bf k},\sigma_{\perp}) \neq E(-{\bf k},\sigma_{\perp}),
\label{eq1}
\end{equation}  
and time-reversal symmetry: 
\begin{equation}
E({\bf k},\sigma_{\perp})=E(-{\bf k},-\sigma_{\perp}).
\label{eq2}
\end{equation} 
Here $\sigma_{\perp}$ is the spin polarization and {\bf k} is the momentum. Remarkably, the spin symmetry in the momentum space, $[C_{2\perp}||E]$ ($E$ is the real-space identity), protects a momentum-independent spin-polarization $\sigma_{\perp}$ along the $C_{2\perp}$ rotation axis. In other words, the coplanar spin arrangement on the crystal enforces a collinear spin polarization in the momentum space perpendicular to the spins on the crystal. This starkly contrasts the spin-quantization axis in the spin-polarized band structure of collinear ferromagnets and altermagnets, which is oriented along the magnetic moments of the crystal. 
Finally, the position of momentum space nodal surfaces with a vanishing spin polarization can be constrained by symmetries combining spin and crystal space rotations (proper or improper).

In Fig.~1(C), we show 2D model of a p-wave magnetic lattice that fulfils the above specified spin symmetry criteria. We highlight the opposite spin sublattices related by the $[C_{2\perp}||\boldsymbol{t}]$ symmetry by dark blue and purple arrows in Fig.~1(C). We consider an effective minimal Hamiltonian:
\begin{equation}
H=2t(\cos \frac{k_{x}}{2}\tau_{1}+\cos k_{y})+2t_{J}(\sin \frac{k_{x}}{2}\sigma_{1}\tau_{2}+\cos k_{y} \sigma_{2}\tau_{3}).
\end{equation}
Here, the Pauli matrices $\sigma$ and $\tau$ correspond to the spin and site degree of freedom, respectively, and we consider electron hopping among the grey nonmagnetic sites parameterized by $t$. The p-wave spin splitting arises from the additional exchange-dependent hopping parametrized by $t_{J}$. This hopping modulation captures the exchange field from the magnetic atoms felt by the electron hopping among the grey nonmagnetic sites. We set $t_{J}=0.25t$. 
The magnetic ordering doubles the size of the nonmagnetic unit cell of the square lattice and exhibits the $[C_{2\perp}||\boldsymbol{t}]$ symmetry. The corresponding magnetic unit cell, outlined by the black box in Fig.~1(C), includes two nonmagnetic atoms.

\begin{figure*}[t!]
	\centering
	\includegraphics[width=\linewidth]{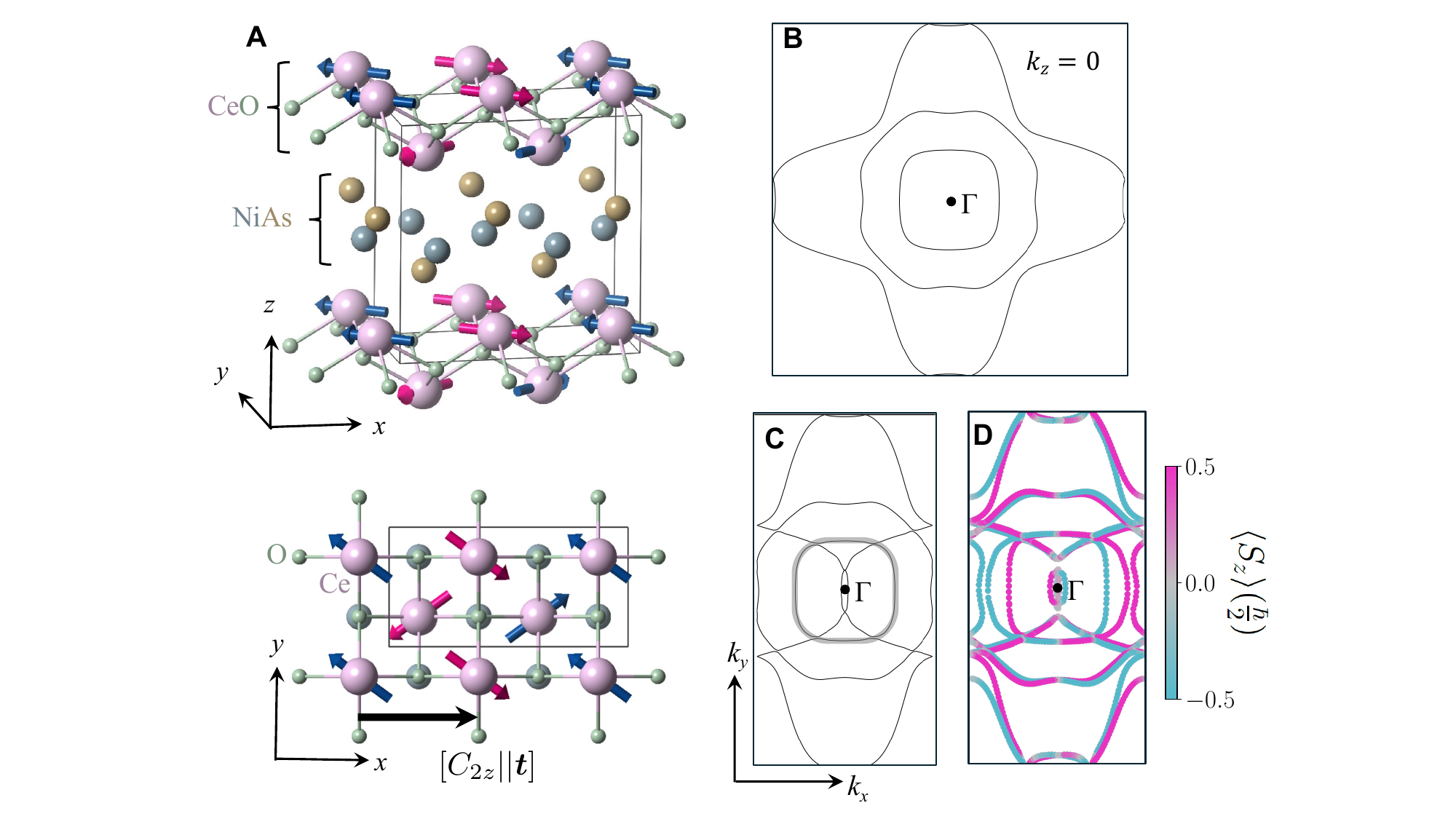}
	\caption{\textbf{Unconventional p-wave magnet CeNiAsO.}
	(A) Side and top view of the magnetic and crystal structure of CeNiAsO with the spin symmetry $[C_{2z}||\boldsymbol{t}]$. (B) DFT Fermi surface in the momentum-space $k_{z}=0$ plane in the non-magnetic crystal with  the in-plane four-fold symmetry. (C) Same as (B) for a crystal unit cell doubled along the $x$-axis to facilitate a direct comparison to the magnetically ordered phase, shown in (D). Magenta/cyan energy iso-surfaces in (D) correspond to opposite spin projections along the spin-rotation axis $C_{2z}$. 
	}
	\label{fig:figure2}
\end{figure*}
  
In Fig.~1(B) and (D) we show the model Fermi surface and band structure, which fulfils Eqs.\eqref{eq1} and \eqref{eq2}, and we show the p-wave modulation of the Fermi surface with single spin degenerate nodal line (marked by grey line $k_{x}=0$ in Fig.~1(B)), a linear band crossing around the $\Gamma$-point in Fig.~1(D), and a spin-polarization in momentum space perpendicular to the direct space exchange fields. The magenta and cyan colouring in panels B and D correspond to the positive and negative signs of the spin projection, respectively. The full energy range band structure with the projected spin polarization is shown in Supplementary Materials Fig.~1.

We now demonstrate, based on our symmetry guidelines and DFT calculations, a realistic material candidate for unconventional p-wave magnetism. We calculate the band structure using the Vienna Ab Initio Simulation Package (VASP), with the Perdew-Burke-Ernzerhof (PBE) exchange-correlation functionals \cite{Kresse1999,Blochl1994,Perdew1996}, constrained non-collinear
magnetism, symmetrization switched off, and in the non-relativistic limit (without relativistic spin-orbit coupling), see Methods.

In Fig.~\ref{fig:figure2}(A), we show the magnetic crystal structure of CeNiAsO reported experimentally \cite{Wu2019,Luo2014}. The non-magnetic unit cell contains two formula units. The crystal structure can be seen as an alternating stacking of quasi-2D layers of Ce-O and Ni-As, with both layers taking the PbO crystal structure (the same crystal structure as FeSe \cite{Mazin2023a}). The corresponding space group is $P4/nmm (\#129)$.

The spin space group of the magnetic crystal is a direct product of the spin-only group and the nontrivial spin-space group \cite{Litvin1974,Litvin1977,Smejkal2021a,Liu2021}, $\boldsymbol{r}_{s} \times \boldsymbol{G}^{S}$. For the experimentally reported coplanar non-collinear magnetic order of CeNiAsO, $\boldsymbol{r}_{s} =\lbrace{E,\bar{C}_{2z}\rbrace}$, where $\bar{C}_{2z}$ is ${C}_{2z}$ combined with spin-space inversion (time reversal), and the spin-rotation $z$-axis is orthogonal to the spin-space $x-y$ plane of the coplanar spin arrangement. 

\begin{figure*}[t!]
	\centering
	\includegraphics[width=\linewidth]{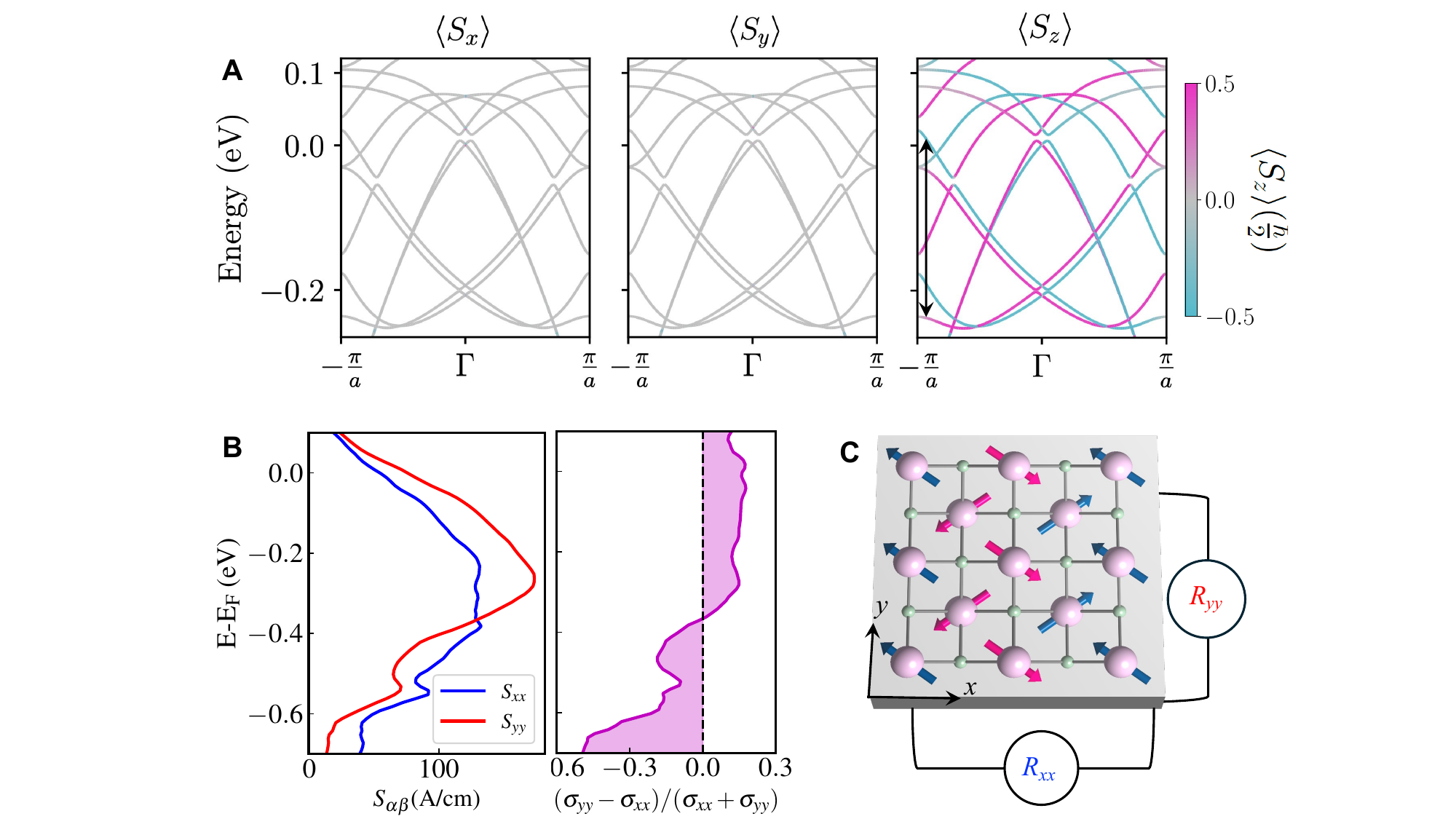}
	\caption{\textbf{Band structure and resistive anisotropy of p-wave magnet CeNiAsO.}
	(A) DFT spin-unpolarized energy bands for spin projections along $x$ and $y$ axes of the spin space, and spin-polarized energy bands for the spin projections along the $z$-axes of the spin space, plotted along the $-X\Gamma X$ path in the momentum space. The spin splitting magnitude  reaches $\sim$200 meV which is marked by the double arrow. (B) DFT longitudinal conductivity along $x$ and $y$ crystal axes.  In the non-magnetic phase, the $x$ and $y$ crystal directions are equivalent, however, the unconventional p-wave magnetism breaks this symmetry, as highlighted on the plotted anisotropy of the conductivity. (C) Proposed experimental geometry for measuring the spontaneous resistivity anisotropy.
	}
	\label{fig:figure3}
\end{figure*}

Following the algorithm for determining the nontrivial spin-space group \cite{Smolyanyuk2024i,Shinohara2024}, we find that $\boldsymbol{G}^{S}$ in CeNiAsO contains the following symmetry elements:
$[E||E]$, $[E||M_{y}\boldsymbol{t}_{0.5b}]$, $[C_{2z}||\boldsymbol{t}_{0.5a}]$, $[C_{2z}||M_{y} \boldsymbol{t}_{0.5a+0.5b}]$, $[C_{2x}||C_{2y}\boldsymbol{t}_{0.5b}]$, $[C_{2y}||C_{2y} \boldsymbol{t}_{0.5a+0.5b}]$, $[C_{2x}||P]$, $[C_{2y}||P \boldsymbol{t}_{0.5a}]$, 
where $C$, and $M$ denote rotation and mirror operations, and $\boldsymbol{t}_{0.5a+0.5b}$ marks lattice translation by a vector $(\frac{1}{2}a,\frac{1}{2}b,0)$ where $a$ and $b$ are lattice parameters. We see that $\boldsymbol{G}^{S}$ contains the $[C_{2z}||\boldsymbol{t}_{0.5a}]$ symmetry. This in combination with the coplanar non-collinear spin-only group, and with broken inversion symmetry $[E\vert\vert P]$, fulfils the above symmetry conditions for the unconventional odd-parity-wave magnetic state. Additionally, the existence of the p-wave state with a single spin-unpolarized line $k_{x}=k_{z}=0$ in the $k_{z}=0$ plane is enforced by the $[C_{2y}||C_{2y}]$ symmetry, implying $E(k_x , k_y , k_z ,\sigma_{z} )=E(-k_{x} , k_{y} , -k_{z} ,-\sigma_{z})$.

We point out that $\boldsymbol{G}^{S}$ of CeNiAsO does not contain any symmetry element with the fourfold real-space rotation symmetry of the non-magnetic crystal. This is a consequence of the antiparallel alignment of spins on atoms along the crystal $x$-axis and the parallel alignment along the $y$-axis. Apart from the the p-wave parity-breaking, the spin group of CeNiAsO thus also indicates an additional anisotropic distortion of the spin-polarized Fermi surfaces.

All these expectations from the spin-group symmetries of CeNiAsO in direct space are confirmed by the DFT calculations of the momentum-space Fermi surfaces, shown in Figs.~2(B-D). Our Fermi surfaces calculated in the nonmagnetic state are shown in Fig.~2(B) and are consistent with a previous report \cite{Wu2019}. While in the non-magnetic state, the spin-unpolarized Fermi surface respects the in-plane four-fold symmetry of the crystal lattice, the spin-polarized Fermi surfaces in  the magnetic state break the parity-symmetry and, in addition, lower the rotation symmetry of the crystal from four-fold to two-fold. Also consistent with the spin-group symmetries, the momentum-independent spin-polarization axis is along the spin-space $z$-axis.

In Fig.~\ref{fig:figure3}(A), we show the DFT band structure. As expected, the spin-polarization is observed only for the spin-projection along the $z$-axis. The bands spin-split linearly in momentum near the $\Gamma$-point, in concurrence with the parity breaking, and the splitting reaches a 200~meV scale, highlighted by the double arrows in Fig. 3(A). This extraordinarily large spin splitting in a time-reversal symmetric band structure is due to its non-relativistic exchange origin. 

Fig.~\ref{fig:figure3}(B) shows DFT calculations of the longitudinal conductivity and spontaneous resistive anisotropy of CeNiAsO in the magnetic phase (see Methods). As mentioned above, the non-magnetic crystal lattice of CeNiAsO exhibits the space group $P4/nmm (\#129)$. Due to the four-fold symmetry, the longitudinal conductivity in the $x-y$ plane is isotropic in the non-magnetic state. The unconventional p-wave magnetic phase spontaneously breaks this symmetry, resulting in a large resistive anisotropy, as shown in Fig.~\ref{fig:figure3}(B). 
The transition from isotropic to anisotropic resistivity can thus serve as a signature, directly accessible by electrical measurements (Fig.~\ref{fig:figure3}(C)), of the transition to the unconventional symmetry-breaking phase of the p-wave magnet CeNiAsO.

According to our symmetry guidelines, we have identified in the Magndata database \cite{Gallego2016}  73 material candidates for the unconventional odd-parity-wave magnetism featuring the coplanar non-centrosymmetric magnetic crystal structure with the $[C_{2\perp}||\boldsymbol{t}]$ symmetry. They are listed in Supplementary Materials Tab.~1. Furthermore, we derived spin point groups of all the materials, which allows us to sort the material candidates into 43 p-wave magnets with one spin unpolarized surface in the momentum space, and 30 f-wave magnets with three spin unpolarized surfaces. In the table, we list also the crystallographic orientation of the induced p-wave spin polarization to guide experiments. 

Our identification of p-wave magnetism, based on the non-relativistic spin symmetries of crystals with a non-collinear magnetic order, is conceptually distinct from previous attempts to identify unconventional p-wave ordering via strongly correlated Fermi-liquid instabilities \cite{Born1948,Bohm1949,Hirsch1990,Wu2007,Jung2015,Kiselev2017,Wu2018} or hidden orders \cite{Knafo2020}. Unlike correlated metallic electron fluids \cite{Fradkin2010}, and alike altermagnetism, 
our spin-symmetry based approach allows p-wave magnetism to materialize even at the single quasi-particle level of DFT. 
As a result, our p-wave magnetism opens the possibility of its robust material realization, direct observation by photoemission spectroscopy, and device exploitation in metallic and insulating phases. 

To conclude, we have identified the analogy of superfluid $^{3}$He in unconventional p-wave magnetism, which has been elusive for many decades.  While in our unconventional p-wave magnetism we consider the ordering of electrons in the charge and spin channels, in superfluidity \cite{Leggett1975} the ordering corresponds to the p-wave Cooper pairing of the fermionic $^{3}$He atoms \cite{Vollhardt2013}. The superfluid $^{3}$He is known to host a plethora of rich and anisotropic properties. In turn, similarly intriguing physics can now emerge in p-wave magnets. P-wave spin currents are an example of extraordinary phenomena anticipated in p-wave magnets \cite{Wu2004}. Attractive properties of p-wave magnets are the giant magnitude of odd-parity wave exchange spin splittings, strongly anisotropic band structure, collinear p-wave spin polarization and the possibility to control the spin-polarization by manipulating the direct-space magnetic order. The linear p-wave dispersion around the $\Gamma$-point akin to relativistic spin-orbit coupled band structures is also a promising feature to be explored in the context of topological quasi-particles \cite{Smejkal2018,AndreiBernevig2022}.

\section*{Acknowledgments}

We acknowledge fruitful discussions with Jörg Schmalian, Rafael M. Fernandes, Andrey Chubukov, Rafael González Hernández, Venkata Krishna Bharadwaj, and Nayra Alvarez. We acknowledge funding from the Czech Science Foundation Grant No. 19-18623X,  the Ministry of Education of the Czech Republic Grants No. LM2018096, LM2018110, LM2018140, and LNSM-LNSpin, CZ.02.01.01/00/22008/0004594, ERC Advanced Grant no. 101095925, Deutsche Forschungsgemeinschaft Grant TRR 173 268565370 (project A03) and TRR 288 42221347 (project B05), the Johannes Gutenberg University Grant TopDyn, and  the computing time granted on the supercomputer Mogon at Johannes Gutenberg University Mainz (hpc.uni-mainz.de).

\section{Methods}

\textbf{Density functional theory.} 
We constrained the magnetic moment direction and size using VASP, with symmetrization and spin-orbit coupling turned off.
The constraining procedure introduces an energy penalty if the magnetic moments deviate from the constraining magnetic moment \cite{VASP_wiki_constrained_magnetic_moment}.
The lattice parameters for CeNiAsO are $a = 8.124\,\mathrm{\AA}$, $b =4.062\,\mathrm{\AA}$, and $c = 8.106\,\mathrm{\AA}$ for the p-wave magnetic phase below $7.6\,K$ \cite{Wu2019}.
We obtained the best results in terms of total energy, the penalty energy, and the size of the local magnetic moments when setting the magnetic atom integration radius as large as possible. 
We used integration radii $1.98\,\mathrm{\AA}$ for Ce, $1.21\,\mathrm{\AA}$ for Ni, $1.09\,\mathrm{\AA}$ for As, and $0.24\,\mathrm{\AA}$ for O, filling $98\%$ of the smallest Ce-Ce distance, $94\%$ of the Ce-O distance, and $98\%$ of the Ni-As distance.
The latter three were based on covalent radii.
For this choice, the penalty energy was $2.77\times 10^{-8}\,\mathrm{eV}$ with a one-shot self-consistency calculation. 
For this calculation, we used $\lambda=2\,\mathrm{eV}/\mu_{\mathrm{B}}^2$, constraining magnetic moments ($(\pm 0.33,\pm 0.22,0)\,\mu_{\mathrm{B}}$) on the Ce atoms corresponding to experimentally obtained values below $7.6\,K$ \cite{Wu2019}, and the initial magnetic moments being twice as large.
We used a $4\times 8\times 4$ k-grid and a $460\,\mathrm{eV}$ energy cut-off.
We converged every \textit{ab initio} calculation within $5\times10^{-8}\,\mathrm{eV}$.

We plotted the spin polarized energy bands and Fermi surfaces using PyProcar version 5.6.6 \cite{Vanderbilt2018}.

In this section we provide details on calculating the longitudinal Ohmic conductivity of the  p-wave magnet CeNiAsO. We have constructed tight-binding model Hamiltonian of CeNiAsO by deploying atom-centered Wannier functions within the Vasp2Wannier90 code\cite{Marzari1997} and we have calculated the Ohmic conductivity as implemented in postW90 code\cite{Pizzi2020}. 
The longitudinal conductivity ($\sigma$) can obtained as,
\begin{equation}
    \sigma_{ij}= (2\pi e\tau/\hbar) S_{ij}
\end{equation}
with, \begin{equation}
    S_{ij}=\frac{e}{h}\int  \sum_n \frac{\delta E_n}{\delta k_i} \frac{\delta E_n}{\delta k_j}\left(-\frac{\delta f_0}{\delta E}\right)_{E=E_n} d\mathbf{k}~.
\end{equation}
We numerically evaluate the integral by employing fine momentum space mesh $101\times 201\times 101$. 

Finally, we define spontaneous resistive anisotropy ratio as,
\begin{equation}
\frac{\rho_{xx}-\rho_{yy}}{\rho_{xx}+\rho_{yy}}=\frac{\sigma_{yy}-\sigma_{xx}}{\sigma_{xx}+\sigma_{yy}}=\frac{S_{yy}-S_{xx}}{S_{xx}+S_{yy}}.
\end{equation}
The spontaneous resistive anisotropy ratio does not depend on the relaxation time $\tau$.

\newpage

\end{document}